\documentclass[prb,twocolumn]{revtex4}
\usepackage{graphicx}
\usepackage{dcolumn}
\usepackage{amsmath,amsbsy,amssymb,graphics}
\usepackage{bm}
\usepackage{mathrsfs,color}

\newcommand{\diff}{\mathrm{d}}
\newcommand{\imag}{\mathrm{Im}\,}

\newcommand{\imu}{\mathrm{i}}
\newcommand{\epn}{\mathrm{e}}

\newcommand{\ua}{\uparrow}
\newcommand{\da}{\downarrow}
\newcommand{\dg}{\dagger}
\newcommand{\la}{\langle}
\newcommand{\ra}{\rangle}
\newcommand{\al}{\alpha}
\newcommand{\sg}{\sigma}
\newcommand{\gm}{\gamma}
\newcommand{\ep}{\varepsilon}

\begin{document}

\preprint{APS/123-QED}

\title{
Mean-field description of odd-frequency superconductivity 
\\
with staggered ordering vector
}

\author{Shintaro Hoshino}

\affiliation{
Department of Basic Science, The University of Tokyo, Meguro, Tokyo 153-8902, Japan
}

\date{\today}

\begin{abstract}
A low-energy fixed-point Hamiltonian is constructed for the $s$-wave odd-frequency pairing state with staggered ordering vector in the two-channel Kondo lattice.
The effective model is justified because it reproduces low-energy behaviors of self energy obtained by the dynamical mean-field theory.
The retardation effect is essential for the odd-frequency pairing, which comes from the hybridization process between conduction electrons and pseudofermions originating from localized spins at low energies.
Using the effective Hamiltonian, the electromagnetic response functions are microscopically calculated.
The present system shows the ``weak'' Meissner effect, where both paramagnetic and diamagnetic parts contribute to the Meissner kernel to give a small total diamagnetic response in the superconducting state.
This feature is in contrast to the ordinary $s$-wave BCS pairing where only the diamagnetic kernel is finite in the ground state.
The staggered nature of the odd-frequency order parameter plays an important role for the sign of the Meissner kernel.
\end{abstract}

\pacs{Valid PACS appear here}
\maketitle

\section{Introduction}

Diversity of superconducting phenomena has been attracting continued attention since the discovery of unconventional superconductivity in a variety of strongly correlated systems.
The superconducting state of matter is characterized by Cooper pairs, whose properties are classified by space-time and spin structures.
Focusing on the time structure of the particle pair, we can have odd-frequency (OF) superconductivities in addition to ordinary even-frequency (EF) pairing, which has extended the concept of unconventional superconductivity \cite{berezinskii74, balatsky92}.
It is also known that the OF pairing state can alternatively be interpreted as an EF composite pairing state, where the order parameter is described by two-body quantities \cite{emery92, balatsky93}.
In this picture we no longer have to consider the time dependence of order parameters.

Theoretically possible realizations of the OF superconductivity in bulk systems are proposed in electron systems such as 
$t$-$J$ model \cite{balatsky93}, 
Kondo and Anderson lattices \cite{emery93, coleman93, coleman94, coleman95, zachar96, jarrell97, coleman97, coleman99, anders02, anders02-2, flint11, hoshino14}, 
geometrically frustrated system \cite{vojta99}, 
quantum critical regime \cite{fuseya03, yada08, hotta09}, 
quasi-one dimensional systems \cite{shigeta09, shigeta11, yanagi12, shigeta13}, 
and electron-phonon coupled systems \cite{abrahams93, kusunose11}.
However, using mean-field type theory, it has been shown that the second-order phase transition into the OF superconductivity is unstable in general \cite{heid95, kusunose11-2}.
In this case the Meissner kernel has the sign opposite to that in the ordinary BCS theory.
Such unphysical result is called the ``negative Meissner effect''.
On the other hand, the OF superconductivity has also been investigated in non-uniform systems such as superconductor/ferromagnet heterostructures, where the OF pairing state appears by a proximity effect \cite{bergeret05, tanaka07, tanaka07-2, asano07, eschrig07, yokoyama11,tanaka12}.
In this case, unlike bulk systems, we do not have thermodynamic difficulty to realize the OF superconducting state.

In order to resolve the thermodynamic instability in bulk systems, some ideas have been put forward: the OF superconductivity is possibly stabilized by considering first-order transitions instead of second-order ones, strong-coupling corrections beyond simple mean-field theory, or spatially inhomogeneous pairing amplitude \cite{coleman94, heid95}.
From another point of view, it is also discussed within path-integral approach that the homogeneous OF superconducting state cannot be stable in Hermitian mean-field Hamiltonian but can exist if one assumes the non-Hermitian relations for anomalous Green functions \cite{belitz99, solenov09, kusunose11-2}.
Another study has demonstrated that the positive Meissner effect is found using the composite-operator description \cite{abrahams95, dahal09}.
Whereas the several proposals have thus been submitted, the necessary conditions for the stable OF superconductivity are still unclear at present.

To get further insights on bulk OF pairing states, it is useful to focus on a microscopically established model that shows the OF superconductivity.
Recently we have demonstrated that the $s$-wave OF superconductivity with staggered ordering vector is stable in the two-channel Kondo lattice (TCKL), by using the dynamical mean-field theory (DMFT) which takes full account of local correlations \cite{hoshino14}.
In this paper, we take the TCKL as a concrete example of the OF pairing, and address the following two issues; i) possibility of mean-field description for OF superconductivity and ii) sign of the Meissner kernel in the pairing state.

We will show that the staggered $s$-wave OF pairing in the TCKL can be described by a Hermitian mean-field Hamiltonian.
This state is classified into a different category from the one proposed in the previous studies \cite{solenov09,kusunose11-2}.
The constructed effective one-body model illuminates the pairing mechanism of this exotic superconductivity, where the Cooper pair is formed through the hybridization of conduction electrons with localized pseudofermions.
We microscopically calculate the electromagnetic response function in the introduced model on the tight-binding lattice, and demonstrate that the ordinary positive Meissner effect is obtained in the staggered OF superconducting state.

This paper is organized as follows.
In the next section, we introduce the TCKL and define the unitary transformations which are useful to discuss composite ordered states.
Section III is devoted to the construction of effective one-body model that describes the low-energy behaviors of the OF superconducting state.
The electromagnetic responses in the TCKL are calculated in Sect. IV.
We discuss characteristics of the present superconductivity in Sect. V, and summarize the results in Sect. VI.

\section{Two-Channel Kondo Lattice and Composite Orders}

Let us begin with the TCKL Hamiltonian \cite{jarrell96}
\begin{align}
{\cal H} = \sum_{\bm k \al \sg} (\ep_{\bm k} - \mu) c^\dg_{\bm k\al\sg} c_{\bm k \al\sg}
+ J \sum_{i\al} \bm S_i \cdot \bm s_{{\rm c}i\al}
, \label{eq:hamilt}
\end{align}
where the channel degree of freedom is labeled by $\al=1,2$, and pseudospin by $\sg=\ua,\da$.
The operators $\bm S_i$ and $\bm s_{{\rm c}i\al} = \frac 1 2 \sum_{\sg\sg'} c_{i\al\sg}^\dg \bm \sg_{\sg\sg'} c_{i\al\sg'}$ are the localized spin and conduction-electron spin at site $i$, respectively, with $\bm \sg$ being the Pauli matrix.
This model has the SU(2) symmetry both in the channel and spin spaces, which are separately denoted as SU(2)$_{\rm C}$ and SU(2)$_{\rm S}$.
We take the three-dimensional cubic lattice for simplicity.
The kinetic energy of conduction electrons is then given by
\begin{align}
\ep_{\bm k} = -2t \sum_{\nu=x,y,z} \cos k_\nu
, \label{eq:disp_3d}
\end{align}
where the lattice constant is taken as unity.
This satisfies the relation $\ep_{\bm k} + \ep_{\bm k+\bm Q} = 0$ with $\bm Q = (\pi,\pi,\pi)$ being the staggered ordering vector.
We take $t=1$ as a unit of energy in the following.

Physically the TCKL describes non-Kramers doublet systems with $f^2$ configuration per site as realized in Pr- and U-based systems \cite{cox98}.
Here the channel $\alpha$ and pseudospin $\sigma$ are regarded as real spin and orbital degrees of freedom, respectively.
The channel (real spin) degeneracy in this case is protected by the time-reversal symmetry.
Hereafter we simply call $\alpha$ and $\sigma$ ``channel'' and ``spin'', respectively.
When we discuss the relevance to real non-Kramers systems, we should be careful about the correspondence between the index in the TCKL and physical degrees of freedom.

We will describe the OF superconducting state in the TCKL by constructing effective mean-field theory in this paper.
To this end, we first briefly summarize the known properties of the TCKL.
According to the previous studies \cite{jarrell96, hoshino13}, the system shows a non-Fermi liquid behavior in the disordered phase even at low temperatures.
This is due to the residual entropy that cannot be removed by the Kondo effect in the two-channel case.
Hence the ground state must resolve this entropy by some phase transition.
The DMFT study has demonstrated that the TCKL indeed shows electronic orderings at low temperatures \cite{hoshino11, hoshino14}.
Among them, we especially concentrate on the composite orders, which are characterized by the order parameter involving both the conduction electrons and localized spins.
These ordered states can also be regarded as the OF orders.

We have two kinds of composite orders in the TCKL: channel and superconducting orders.
It is convenient to deal with these two orders on equal footing.
The operators for composite order parameters are written as
\begin{align}
\bm \Psi &= \sum_{i\al\al'\sg\sg'} c^\dg_{i\al\sg} \bm \sg_{\al\al'} (\bm S_i \cdot \bm \sg_{\sg\sg'}) c_{i\al'\sg'}
, \label{eq:composite_channel}
\\
\Phi^+ &= \sum_{i\al\al'\sg\sg'} c^\dg_{i\al\sg} \epsilon_{\al\al'} [\bm S_i \cdot (\bm \sg\epsilon)_{\sg\sg'}] c^\dg_{i\al'\sg'} \epn^{\imu \bm Q \cdot \bm R_i}
, \label{eq:composite_super}
\end{align}
where $\epsilon = \imu \sg^y$ is the antisymmetric unit tensor and $\bm R_i$ is a spatial coordinate of the site $i$.
We also define the operators by $\Psi^{\pm} = \Psi^x \pm \imu \Psi^y$ and $\Phi^{\pm} = \Phi^x \pm \imu \Phi^y = (\Phi^{\mp})^\dg$.
Here the finite $\la \bm \Psi \ra$ corresponds to the channel SU(2)$_{\rm C}$ symmetry breaking, while the order parameter $\la \Phi^\pm \ra$ breaks the gauge U(1) symmetry.
The form factors $\epn^{\imu \bm Q \cdot \bm R_i}$ and $\epsilon_{\al\al'}\epsilon_{\sg\sg'}$ in Eq.~\eqref{eq:composite_super} represent the staggered ordering vector and channel-singlet/spin-singlet pairing, respectively.

\begin{figure}[t]
\begin{center}
\includegraphics[width=65mm]{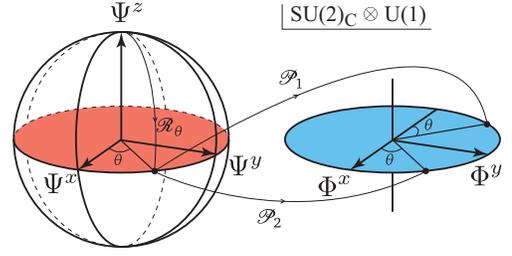}
\caption{
(Color online)
Unitary transformations of the operators defined by Eqs.~\eqref{eq:composite_channel} and \eqref{eq:composite_super}.
The left and right panels illustrates the channel SU(2)$_{\rm C}$ and gauge U(1) spaces.
The channel rotation $\mathscr{R}_\theta$ does not change Eq.~\eqref{eq:hamilt}, while the Hamiltonian is invariant under the particle-hole transformation $\mathscr{P}_\al$ only at half filling.
}
\label{fig:so5}
\end{center}
\end{figure}

The above operators are related to each other by the unitary transformations.
One of them is the rotation in channel space from $z$-axis onto $xy$-plane defined by
\begin{align}
\mathscr{R}_\theta c_{\bm k \al\sg} \mathscr{R}_\theta^{-1} &= [\epn^{\imu\theta/2} c_{\bm k 1\sg} +\sg^z_{\al\al} \epn^{-\imu\theta/2} c_{\bm k 2\sg} ] / \sqrt 2
, \label{eq:channel_rotation}
\end{align}
where the angle $\theta$ is measured from $x$-axis.
On the other hand, the charge-conjugation transformation for channel $\al$ is defined by
\begin{align}
\mathscr{P}_\al c_{\bm k \al\sg} \mathscr{P}_\al^{-1} &= \sum_{\sg'} \epsilon_{\sg\sg'} c^\dg_{-\bm k - \bm Q, \al \sg'}
. \label{eq:charge_conjugation}
\end{align}
These transformations do not change the localized-spin operators.
The operators \eqref{eq:composite_channel} and \eqref{eq:composite_super} are connected with each other by the following relations:
\begin{align}
\mathscr{R}_\theta \Psi^z \mathscr{R}_\theta^{-1}
&= \cos \theta \Psi^x + \sin \theta \Psi^y
, \label{eq:transf_ex1} \\
\mathscr{P}_1 \Psi^{\pm} \mathscr{P}_1^{-1} 
&= - \Phi^{\mp}
, \label{eq:transf_ex2} \\
\mathscr{P}_2 \Psi^{\pm} \mathscr{P}_2^{-1} 
&= \Phi^{\pm}
. \label{eq:transf_ex3}
\end{align}
This is schematically drawn in Fig.~\ref{fig:so5}.
These unitary transformations can also be written in terms of ten SO(5) generators for conduction electrons \cite{demler04}.

At half filling with $\mu=0$, the TCKL Hamiltonian \eqref{eq:hamilt} is invariant under both the transformations $\mathscr{R}_\theta$ and $\mathscr{P}_\al$, which is the consequence of the SO(5) symmetry \cite{affleck92, hattori12}.
Hence the ordered states described by $\bm \Psi$ and $\Phi^\pm$ are degenerate at $\mu=0$.
Away from half filling, on the other hand, the symmetry is lowered into SU(2)$_{\rm C}$$\otimes$U(1).
In this case the composite pairing state with $\Phi^\pm$ is more stable than the composite channel order as demonstrated by numerical calculations \cite{hoshino14}.

One can also consider another form of the order parameter that includes only conduction electrons.
The corresponding operator is given by \cite{hoshino14}
\begin{align}
\bm \psi_{\rm c} &=
\sum_{\bm k\al\al'\sg} \ep_{\bm k} c^\dg_{\bm k\al\sg}\bm\sg_{\al\al'} c_{\bm k\al'\sg}
\label{eq:order_prm_cond1}
, \\
\phi_{\rm c}^+ &=  \sum_{\bm k\al\al'\sg\sg'} \ep_{\bm k} c^\dg_{\bm k\al\sg}
\epsilon_{\al\al'} \epsilon_{\sg\sg'} c^\dg_{-\bm k-\bm Q,\al'\sg'}
\label{eq:order_prm_cond2}
.
\end{align}
The expectation values of Eqs.~\eqref{eq:order_prm_cond1} and \eqref{eq:order_prm_cond2} become finite at the same time as the composite order parameters appear.
If we rewrite them in terms of the real-space basis, we find that these are composed of non-local quantities with nearest neighbor sites.
However, we can have this order parameter even when the self energy is local as in the DMFT framework \cite{hoshino11}.
Hence these order parameters are interpreted as induced from the local composite order parameter secondarily.
We note that the form factor $\ep_{\bm k}$ in Eq.~\eqref{eq:order_prm_cond2} 
is essential, while it becomes identically zero if we do not have this factor.

We comment on the relation between the OF order parameters and composite quantities for superconductivity.
The OF pairing amplitude with channel-singlet (Cs) and spin-singlet (Ss) is defined by
\begin{align}
O^+_{\rm CsSs} (\bm q, \tau) = \sum_{i\al\al'\sg\sg'} c^\dg_{i\al\sg} (\tau) 
\epsilon_{\al\al'} \epsilon_{\sg\sg'} c^\dg_{i\al'\sg'}
\epn^{-\imu\bm q\cdot \bm R_i}
.  \label{eq:odd_op}
\end{align}
This quantity is an odd-function with respect to imaginary time: $\la T_\tau O^+_{\rm CsSs} (\bm q, -\tau) \ra = -\la T_\tau O^+_{\rm CsSs} (\bm q, \tau) \ra$.
Therefore the value at $\tau=0$ is zero unlike ordinary EF pairing states.
On the other hand, the first-order time-derivative of the operator \eqref{eq:odd_op} is finite, which is given by
\begin{align}
\partial_\tau O^+_{\rm CsSs} (\bm Q, \tau)|_{\tau=0} = \phi_{\rm c}^+ + \frac J 2  \Phi^+
 \label{eq:odd_deriv_composite}
\end{align}
at $\bm q= \bm Q$.
Here we do not have to consider the time dependence of the order parameters.
Thus the EF composite order parameter is related to OF pairing amplitude.
This can be applied also to the channel orders.

\section{Effective Low-Energy Hamiltonian}

In this section we derive a low-energy fixed-point Hamiltonian for the composite pairing state with $\Phi^{\pm}$, which reproduces the single-particle properties that are numerically calculated by the DMFT.
For the derivation, we first consider the uniform channel order with $\Psi^z$ and transform it by using symmetry operations.
We then show that the same result can be obtained by a mean-field approximation of the original Kondo interaction.

\subsection{Derivation of one-body Hamiltonian for composite pairing state} \label{sec:eff_one_body}
In the composite channel ordered state with $\Psi^z = 2\sum_i \bm S_i \cdot ( \bm s_{{\rm c}i1} - \bm s_{{\rm c}i2} ) $, the localized spins are coupled selectively with one of the two channels \cite{hoshino11}.
This selective Kondo singlet state can be described by the one-body hybridization model as \cite{hoshino13,chandra13}
\begin{align}
{\cal H}_{\rm eff} &= \sum_{\bm k \al \sg} 
\ep_{\bm k}
c_{\bm k \al \sg}^\dg c_{\bm k \al \sg}
\nonumber \\
&\ \ 
+ \sqrt 2 \sum_{\bm{k}\sg} \left(
V f_{\bm k\sg}^\dg c_{\bm k 1 \sg} + V^* c^\dg_{\bm k 1 \sg} f_{\bm k\sg}
\right)
,
\label{eq:one-body_hamilt}
\end{align}
where the pseudofermion $f_{\bm k\sg}$ is introduced, which effectively describes the localized spins at low energies.
The conduction electrons with $\alpha=2$ are decoupled from localized-spin degrees of freedom.
The hybridization strength $V$ corresponds to the effective mean field for the uniform channel order $\Psi^z$.
The factor $\sqrt 2$ is put for convenience.
The effective one-body Hamiltonian \eqref{eq:one-body_hamilt} is justified because it indeed reproduces the low-energy behavior of the self energy obtained by the DMFT in the lowest order in frequency \cite{hoshino13}.

If the localized pseudofermions hybridize with both channels of conduction electrons with the same strength, the resultant state corresponds to channel ordering in the $\Psi^x$-$\Psi^y$ plane.
The corresponding effective Hamiltonian can be constructed by using the channel rotation $\mathscr{R}_\theta$ as illustrated in Fig.~\ref{fig:so5}.

By performing both the channel rotation and  charge-conjugation transformation on Eq.~\eqref{eq:one-body_hamilt}, we derive the effective one-body model for OF superconductivity in the TCKL.
Noting the relation $[\mathscr{P}_2 \mathscr{R}_\theta, {\cal H}] = 0$ at half filling, we can explicitly write down the Hamiltonian as
\begin{align}
\tilde{\cal H}_{\rm eff} &\equiv
\mathscr{P}_2 \mathscr{R}_\theta {\cal H}_{\rm eff} (\mathscr{P}_2 \mathscr{R}_\theta)^{-1}  
\\
&= \sum_{\bm k \al \sg} \ep_{\bm k} c_{\bm k \al \sg}^\dg c_{\bm k \al \sg}
+ |V| \sum_{\bm{k}\sg\sg'} 
\left[
 \epn^{\imu (\phi + \theta)/2} \delta_{\sg\sg'} f_{\bm k\sg}^\dg c_{\bm k 1 \sg}
 \right.
\nonumber \\
& \hspace{12mm} \left.
+ \,  \epn^{\imu (\phi - \theta)/2} \epsilon_{\sg\sg'} f_{\bm k\sg}^\dg c^\dg_{-\bm k-\bm Q, 2 \sg'}
+ {\rm h.c.}
\right] 
,
\label{eq:one-body_hamilt_tilde}
\end{align}
where we have defined the phase factor $\phi$ by
$V = |V| \epn^{\imu \phi/2}$.
Equation \eqref{eq:one-body_hamilt_tilde} gives the low-energy fixed-point Hamiltonian for the pairing state with the order parameter
$\la  \cos \theta \Phi^x + \sin \theta \Phi^y  \ra$, as seen from Fig.~\ref{fig:so5}.
We note that this effective model at $V=0$ cannot describe a non-Fermi liquid state realized in the disordered TCKL.
Hence the present description is reasonable only in the deep inside of the ordered state.

In Eq.~\eqref{eq:one-body_hamilt_tilde}, the channels $\al=1$ and $\al=2$ seem not symmetric, even though we have the channel-singlet pairing state.
This originates from the fact that the way of introduction of pseudofermions is not unique.
For example, we may replace the operator as
\begin{align}
f_{\bm k\sg} \rightarrow \sum_{\sg'} \epsilon_{\sg\sg'} f_{-\bm k-\bm Q, \sg'}^\dg
,
\end{align}
where the channel indices are interchanged in the resultant Hamiltonian.
Even in this case the physical properties for conduction electrons do not change except for the phase factor.
As shown in the next subsection, we can obtain the symmetric expression between the channels by tracing out the pseudofermion degrees of freedom.
This indicates that localized pseudofermions are nothing but virtual degrees of freedom.
Correspondingly, the phase $\phi$ of the hybridization does not enter in any physical quantities.

Next we construct the wave function for the composite pairing state at half filling.
The diagonalized Hamiltonian is written as
\begin{align}
\tilde{\cal H}_{\rm eff} = \sum_{\bm k \sg}\sum_{p=0,\pm} 
 E_{\bm kp} \gamma^\dg_{\bm k \sg p} \gamma_{\bm k \sg p}
 ,
\end{align}
where the eigenenergies are given by
\begin{align}
E_{\bm k0} &= \ep_{\bm k}
, \label{eq:energy1}
\\
E_{\bm k\pm} &= \frac 1 2 \left( \ep_{\bm k} \pm \sqrt{\ep_{\bm k}^2 + 8|V|^2} \right)
.  \label{eq:energy2}
\end{align}
Namely the present pairing state has the gapless single-particle excitation at the Fermi level even in the pairing state.
The form of the eigenenergies is similar to that obtained in Ref.~\cite{coleman94}.
On the other hand, the Bogoliubov eigenoperators are given by
\begin{align}
\gamma_{\bm k\sg 0} &= \frac{ 1 }{ \sqrt{2} } \left(
\epn^{\imu\theta/2} c_{\bm k1\sg} + \epn^{-\imu\theta/2} \sum_{\sg'} \epsilon_{\sg\sg'} c^\dg_{-\bm k-\bm Q, 2\sg'}
\right)
, \label{eq:bogolon_con} \\
\gamma_{\bm k\sg \pm} &= u_{\bm k\pm} f_{\bm k\sg} + v_{\bm k\pm} \bar \gamma_{\bm k\sg 0}
, \label{eq:bogolon_hyb}
\end{align}
where we have defined the operator
\begin{align}
\bar \gamma_{\bm k\sg 0} &= \frac{ 1 }{ \sqrt{2} } \left(
\epn^{\imu\theta/2} c_{\bm k1\sg} - \epn^{-\imu\theta/2} \sum_{\sg'} \epsilon_{\sg\sg'} c^\dg_{-\bm k-\bm Q, 2\sg'}
\right)
.
\end{align}
The coefficients are given by
\begin{align}
u_{\bm k\pm} &= - \frac{ E_{\bm k\mp} }{ \sqrt{E_{\bm k\mp}^2 + 2|V|^2} }
, \\
v_{\bm k\pm} &= \frac{ \sqrt 2 |V| \epn^{-\imu\phi/2} }{ \sqrt{E_{\bm k\mp}^2 + 2|V|^2} }
.
\end{align}
The coefficients $u_{\bm k\pm}$ and $v_{\bm k \pm}$ represent the weights in the Bogoliubov particles coming from localized pseudofermion and conduction electron, respectively.

Using the eigenoperators, the ground state wave function is simply written as
\begin{align}
|\Phi \ra = 
\prod_{\bm k\in {\rm HBZ}, \sg} \gamma^\dg_{\bm k\sg 0} \prod_{\bm k\sg} \gamma^\dg_{\bm k \sg -}
|\tilde 0\ra
, \label{eq:wave_func_one_body}
\end{align}
where ``HBZ'' means the half Brillouin zone determined by the condition $\ep_{\bm k} < 0$ for the simple cubic lattice.
The introduced state
\begin{align}
|\tilde 0\ra = \prod_{\bm k\sg} c^\dg_{\bm k2\sg} |0\ra
\end{align}
is the vacuum of the Bogoliubov particles, while $|0\ra$ is that of the original fermions.
The former part of Eq.~\eqref{eq:wave_func_one_body} forms the Fermi surface in the pairing state, while the latter part made from the pseudofermions $f_{\bm k\sg}$ and $\bar \gm_{\bm k\sg 0}$ physically corresponds to the Kondo singlet state.

With the effective Hamiltonian, we can also derive the quantity corresponding to the composite order parameter.
Using Eq.~\eqref{eq:odd_op}, we obtain 
\begin{align}
&\partial_\tau O^+_{\rm CsSs} (\bm Q, \tau)|_{\tau=0} = \phi_{\rm c}^+ 
\nonumber \\
&\hspace{6mm} + \frac {|V|} 2 \sum_{i\sg\sg'} \left( 
\delta_{\sg\sg'} c^\dg_{i1\sg} f_{i\sg} + \epsilon_{\sg\sg'} f^\dg_{i\sg} c^\dg_{i2\sg'} \epn^{\imu\bm Q\cdot \bm R_i}
\right)
, \label{eq:odd_deriv_eff}
\end{align}
where we have fixed the phases as $\phi=\theta=0$.
By comparing with Eq.~\eqref{eq:odd_deriv_composite}, the second line in Eq.~\eqref{eq:odd_deriv_eff}, involving both conduction electrons and localized pseudofermions, is the counterpart of $\Phi^+$ in the effective model.

\subsection{Green functions and self energies for conduction electrons}

\begin{figure}[t]
\begin{center}
\includegraphics[width=80mm]{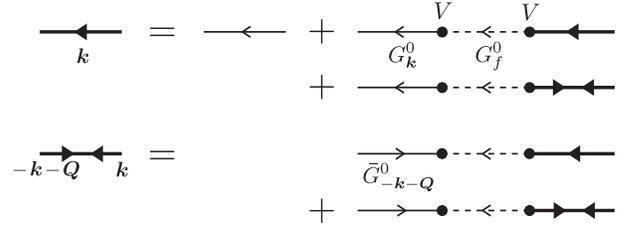}
\caption{
Diagrammatic illustration of the Dyson-Gor'kov equations for conduction-electron Green functions $G$ and $F^\dg$.
The thin and dotted lines show the zero-th order Green functions for conduction electrons and localized pseudofermions, respectively. 
}
\label{fig:EOM}
\end{center}
\end{figure}

Since the localized pseudofermions in the last subsection are artificially introduced in the effective theory, we should trace out these degrees of freedom in order to compare the physical quantities with the original TCKL.
Hence we consider the Green functions only for conduction electrons.
The diagonal and offdiagonal Green functions are defined by
\begin{align}
G_{\bm k\al\sg,\bm k'\al'\sg'}(\tau) &= -\la T_\tau  c_{\bm k\al\sg} (\tau) c^\dg_{\bm k'\al' \sg'}  \ra
, \label{eq_green_def1} \\
F_{\bm k\al\sg,\bm k'\al'\sg'}(\tau) &= -\la T_\tau  c_{\bm k\al\sg} (\tau) c_{\bm k'\al' \sg'}  \ra
, \label{eq_green_def2} \\
F^\dg_{\bm k\al\sg,\bm k'\al'\sg'}(\tau) &= -\la T_\tau  c^\dg_{\bm k\al\sg} (\tau) c^\dg_{\bm k'\al' \sg'}  \ra
, \label{eq_green_def3} \\
\bar G_{\bm k\al\sg,\bm k'\al'\sg'}(\tau) &= -\la T_\tau  c^\dg_{\bm k\al\sg} (\tau) c_{\bm k'\al' \sg'}  \ra
. \label{eq_green_def4} 
\end{align}
The Fourier transformed Green function is defined by
\begin{align}
{\mathscr G} (\imu\ep_n) = \int _0 ^{1/T} {\mathscr G} (\tau) \, \epn^{\imu\ep_n \tau} \diff \tau
\end{align}
where ${\mathscr G}$ represents one of Eqs.~(\ref{eq_green_def1}--\ref{eq_green_def4}), and $\ep_n = (2n+1)\pi T$ is the fermionic Matsubara frequency.
The Green functions are calculated from the equations of motion in the effective one-body Hamiltonian \eqref{eq:one-body_hamilt_tilde}, which are illustrated in Fig.~\ref{fig:EOM}.
Here we have defined the zero-th order Green functions $G^0_{\bm k} (\imu\ep_n) = 1/(\imu\ep_n - \ep_{\bm k})$ for conduction electron and $G^0_f (\imu\ep_n) = 1/\imu\ep_n$ for localized pseudofermion.
By solving the equations, the explicit forms of the Green functions are obtained as
\begin{align}
G_{\bm k\al\sg,\bm k'\al'\sg'} (\imu\ep_n) &= \delta_{\bm k\bm k'} \delta_{\al\al'} \delta_{\sg\sg'}
{\cal G}_{\bm k} (\imu\ep_n)
\nonumber \\
&= - \bar G_{\bm k'\al'\sg', \bm k\al\sg} (-\imu\ep_n)
, \label{eq:green_matsu_g}
\\
F_{\bm k\al\sg,\bm k'\al'\sg'} (\imu\ep_n) &= \delta_{-\bm k-\bm Q, \bm k'} \epsilon_{\al\al'} \epsilon_{\sg\sg'}
{\cal F}_{\bm k} (\imu\ep_n)
\nonumber \\
&= F^\dg_{\bm k'\al'\sg', \bm k\al\sg} (-\imu\ep_n)^*
, \label{eq:green_matsu_f}
\end{align}
where 
\begin{eqnarray}
&\displaystyle 
{\cal G}_{\bm k} (\imu\ep_n) =
\sum_{p=0,\pm} \frac{a_{\bm kp}}{\imu\ep_n - E_{\bm kp}}
, \label{eq:def_g}
&
\\
&\displaystyle 
{\cal F}_{\bm k} (\imu\ep_n) =
\sum_{p=0,\pm} \frac{b_{\bm kp}}{\imu\ep_n - E_{\bm kp}}
, \label{eq:def_f}
&
\\
&a_{\bm k0} = 1/2, \ \ \  a_{\bm k\pm} = |v_{\bm k\pm}|^2 /2,
&
\label{eq:coef_a}
\\
&
b_{\bm k0} = - \epn^{\imu\theta} / 2, \ \ \ b_{\bm k\pm} = \epn^{\imu\theta} |v_{\bm k\pm}|^2 /2
.
&
\label{eq:coef_b}
\end{eqnarray}
We confirm from these expressions that the channels $\al=1$ and $\al=2$ are equivalent.
Note that the phase $\phi$ and the coefficient $u_{\bm k\pm}$, which originate from the localized pseudofermions, do not enter in the conduction electron Green functions.
The diagonal Green function enjoys the translational invariance of the original Hamiltonian, while the offdiagonal one does not.

\begin{figure}[t]
\begin{center}
\includegraphics[width=70mm]{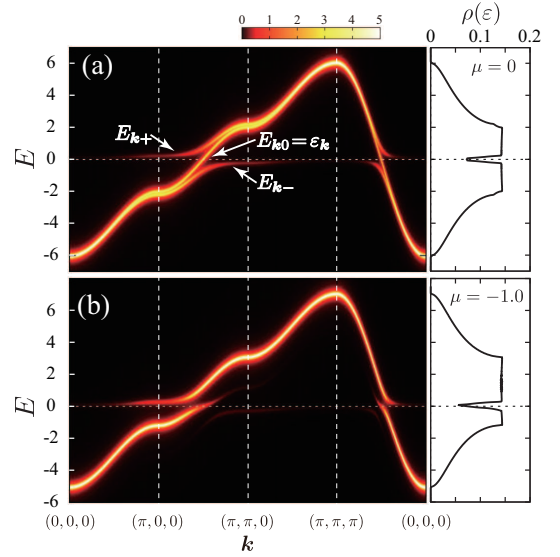}
\caption{
(Color online)
Single-particle spectra $\rho_{\bm k}(\ep)$ for (a) $\mu=0$ and (b) $\mu=-1.0$ in the effective one-body model at $V=0.5$.
Here we have replaced the infinitesimal constant by $\eta = 0.1$.
The right panels show the densities of states $\rho(\ep)$.
}
\label{fig:dos}
\end{center}
\end{figure}

The single-particle spectrum of conduction electrons is defined by
\begin{align}
\rho_{\bm k}(\ep) =  - \frac 1 \pi \imag {\cal G}_{\bm k} (\ep+\imu \eta)
,
\end{align}
where $\eta = +0$.
Figure \ref{fig:dos}(a) shows this angle-resolved spectrum at half filling.
The integrated density of state $\rho(\ep) = N^{-1}\sum_{\bm k}\rho_{\bm k}(\ep)$ is also plotted in the right panel.
As shown in the figure, a half of conduction bands hybridizes with localized pseudofermion, and the others remain decoupled.
These uncoupled bands are different from original conduction electrons, and are composed of the Bogoliubov particles defined by Eq.~\eqref{eq:bogolon_con}.

Now we turn to the dynamical structure of anomalous self energy.
Since the mean-field part of $\tilde {\cal H}_{\rm eff}$ is local, the self energy can also be represented as the local quantity.
To demonstrate this, we introduce the real space-based Green functions by
\begin{align}
G_{ij} &= 
N^{-1}\hspace{-4mm}\sum_{\bm k\bm k'\al\al'\sg\sg'}\hspace{-2mm}
 \delta_{\al\al'}\delta_{\sg\sg'}
G_{\bm k \al \sg, \bm k' \al' \sg'} \, \epn^{\imu (\bm k \cdot \bm R_i - \bm k' \cdot \bm R_j)}
	,
\\
F_{ij} &= 
N^{-1}\hspace{-4mm}\sum_{\bm k\bm k'\al\al'\sg\sg'}\hspace{-2mm}
 \epsilon_{\al\al'}\epsilon_{\sg\sg'}
F_{\bm k \al \sg, \bm k' \al' \sg'} \, \epn^{\imu (\bm k \cdot \bm R_i + \bm k' \cdot \bm R_j)}
,
\end{align}
where we have omitted the Matsubara frequency $\imu\ep_n$.
The self energies are given by the Dyson-Gor'kov equation
\begin{align}
&
\begin{pmatrix}
G_{ij} & F_{ij} \\
F^\dg_{ij} & \bar G_{ij}
\end{pmatrix}
=
\begin{pmatrix}
G^0_{ij} & \\
 & \bar G^0_{ij}
\end{pmatrix}
\nonumber \\
&\hspace{12mm} +
\sum_{\ell}
\begin{pmatrix}
G^0_{i\ell} & \\
 & \bar G^0_{i\ell}
\end{pmatrix}
\begin{pmatrix}
\Sigma_{\ell} & \Delta_{\ell} \\
\Delta^\dg_{\ell} & \bar \Sigma_{\ell}
\end{pmatrix}
\begin{pmatrix}
G_{\ell j} & F_{\ell j} \\
F^\dg_{\ell j} & \bar G_{\ell j}
\end{pmatrix}
.
\end{align}
The free diagonal Green function has been written as $G^0_{ij}$.
The explicit forms of the normal and anomalous self energies are given by
\begin{align}
\Sigma_i (\imu\ep_n) &= \frac{|V|^2}{\imu\ep_n}
, \label{eq:self_ene1} \\
\Delta_i (\imu\ep_n) &= \frac{|V|^2 \epn^{\imu (\theta + \bm Q \cdot \bm R_i)}}{\imu\ep_n}
, \label{eq:self_ene2}
\end{align}
where the site-dependent phase factor $\epn^{\imu(\theta + \bm Q \cdot \bm R_i)}$ represents staggered pairing state.
Equation \eqref{eq:self_ene2} shows that the anomalous self energy, or gap function, is the odd function with respect to frequency.
This gives a clear reason why we regard the present ordered state as the OF superconductivity.

In the effective model, we have the relation \eqref{eq:def_f}, which originates from the Hermiticity of the Hamiltonian.
Hence the superconductivity in the TCKL belongs to a different category from that proposed in the previous studies, where the pairing state cannot be described by any Hermitian mean-field Hamiltonian \cite{belitz99, solenov09, kusunose11}.
We note, however, that the results in this paper do not exclude the possibility of this type of pairing state.

\subsection{Effect of Symmetry-Breaking Fields}

We discuss how much the gapless excitation in the superconducting state is robust against symmetry breaking fields.
In the previous subsections, we have considered the half-filled case that has the SO(5) symmetry.
This high symmetry no longer exists away from half filling due to the finite chemical potential.
The staggered pairing state with $\Phi^{\pm}$ is more stable than the uniform channel order $\bm \Psi$ in this case \cite{hoshino14}.

The effective model for $\mu \neq 0$ is given simply by the Hamiltonian \eqref{eq:one-body_hamilt_tilde} with the chemical potential term added.
The single-particle spectrum at $\mu=-1.0$ is shown in Fig.~\ref{fig:dos}(b).
The finite density of states still remains at the Fermi level, while the Fermi velocity becomes smaller away from half filling.

Other symmetry breaking terms are also examined by adding external fields to Eq.~\eqref{eq:one-body_hamilt_tilde}.
We have confirmed that the single-particle gapless excitation also remains even in the presence of the uniform spin or channel fields.
On the other hand, if we consider the staggered fields, the spectrum can become gapful.
According to the DMFT study in the TCKL \cite{hoshino14}, the staggered spin and channel orders are in the vicinity of the staggered pairing state with $\Phi^\pm$.
Hence if the coexistence phases exist near the phase boundaries, there will be no gapless excitation at the Fermi level.

\subsection{Corresponding Mean-Field Approximation} \label{sect_mfa}
Now we show that the effective one-body model (\ref{eq:one-body_hamilt_tilde}) can be derived from the mean-field approximation of the original TCKL Hamiltonian.
We first write the localized spin in terms of pseudofermions as
\begin{align}
\bm S_i = \frac 1 2 \sum_{\sg\sg'} f^\dg_{i\sg} \bm \sg_{\sg\sg'} f_{i\sg'}
,\label{eq:local_spin_fermi}
\end{align}
with the local constraint
\begin{align}
\sum_{\sg} f^\dg_{i\sg} f_{i\sg} = 1
\label{eq:local_constraint}
\end{align}
for arbitrary site index $i$.
The pseudofermions here are introduced as an equivalent representation of $\bm S_i$, and are not originating from real $f$ electrons itself.
Note that the fermionic representation of localized-spin operators is not unique.
However, if we use Eq.~\eqref{eq:local_spin_fermi}, we can reproduce the results in the previous sections. 
In the mean-field theory for the Kondo lattice, the local constraint \eqref{eq:local_constraint} is satisfied only in the average value \cite{zhang00}, and then we can use the mean-field decoupling.

Let us first consider the composite channel order with $\Psi^z$.
The effective Hamiltonian \eqref{eq:one-body_hamilt} is obtained by decoupling the interaction term as follows:
\begin{align}
J \bm S_i \cdot \bm s_{{\rm c}i1} &\rightarrow 0
, \\
J \bm S_i \cdot \bm s_{{\rm c}i2}  &\rightarrow  \sum_{\sg} \left(
V_2 f^\dg_{i\sg} c_{i2\sg} + V_2^* c^\dg_{i2\sg} f_{i\sg}
\right)
,
\end{align}
where the mean field is defined by
\begin{align}
\frac {3J} 4 \la c^\dg_{i\al\sg} f_{i\sg} \ra = - V_\al
. \label{eq:mean_field1}
\end{align}
We have dropped a constant term.
The resultant Hamiltonian is the same as Eq.~\eqref{eq:one-body_hamilt} if we put $V_2 = \sqrt 2 V$.
The self consistent equation \eqref{eq:mean_field1} in the present mean-field theory is written as
\begin{align}
\frac{4}{3J} = - \frac{1}{N}  \sum_{\bm k} \frac{f(E_{\bm k+}) - f(E_{\bm k-})}{E_{\bm k+} -E_{\bm k-}}
, 
\end{align}
where the Fermi distribution function is defined by $f(x) = 1/ (\epn^{x/T}+1)$.
The equation with $V=0$ determines the transition temperature for $J>0$, and no solution for $J<0$.
This corresponds to the operation of the Kondo effect only for the antiferromagnetic interaction $J>0$.
At half filling, we obtain 
$\rho_0 T_{\rm c} \sim 0.57 \exp [-4/ (3\rho_0 J)]$ for small $J$,
with $\rho_0$ being the bare density of states at the Fermi level.
The condensation energy can also be calculated at zero temperature, which gives 
$\varDelta E = E_{V\neq 0} - E_{V=0} \sim - 2 \rho_0 |V|^2$.
Here the mean field is given by $\rho_0 |V| \sim 0.35 \exp [-2/(3\rho_0 J)]$.
These expressions are the same as those for staggered pairing state discussed in the next, since we have the SO(5) symmetry at half filling.

\begin{figure}[t]
\begin{center}
\includegraphics[width=80mm]{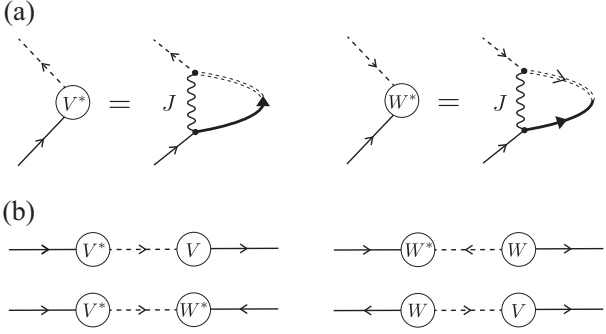}
\caption{
Schematic illustrations for (a) self-consistent equations in mean-field theory and (b) hybridization processes with $V$ and $W$.
The thin and thick lines show the zero-th order and full Green functions, respectively.
}
\label{fig:diagram}
\end{center}
\end{figure}

For the superconductivity corresponding to Eq.~\eqref{eq:one-body_hamilt_tilde}, we decouple the interaction terms as
\begin{align}
J \bm S_i \cdot \bm s_{{\rm c}i1} & \rightarrow  \sum_{\sg} \left(
V_1 f^\dg_{i\sg} c_{i1\sg} + V_1^* c^\dg_{i1\sg} f_{i\sg}
\right)
, \\
J \bm S_i \cdot \bm s_{{\rm c}i2}  & \rightarrow \epn^{\imu \bm Q \cdot \bm R_i} \sum_{\sg\sg'} \epsilon_{\sg\sg'} 
\left(
W_2 f^\dg_{i\sg} c^\dg_{i2\sg'} + W_2^* c_{i2\sg'} f_{i\sg} 
\right)
,
\end{align}
where $V_\al$ is defined by Eq.~\eqref{eq:mean_field1} and the pairing mean-field $W_\al$ by
\begin{align}
\frac {3J} 4 \la c_{i\al\sg} f_{i\sg'} \ra = \epn^{\imu \bm Q\cdot \bm R_i} \epsilon_{\sg\sg'} W_\al
.  \label{eq:mean_field2}
\end{align}
The self-consistent equations \eqref{eq:mean_field1} and \eqref{eq:mean_field2} in the mean-field theory are schematically illustrated in Fig.~\ref{fig:diagram}(a).
If we choose $|V_1| = |W_2| = |V|$, the Hamiltonian is identical to Eq.~\eqref{eq:one-body_hamilt_tilde}, which corresponds to the pure superconductivity.
For $|V_1| \neq |W_2|$, on the other hand, the channel and superconducting orders are mixed.
From the original interaction form, it is difficult to find out this complicated mean-field decoupling for pairing states.
However, it becomes clearer by referring to the effective one-body model derived in Sect.~\ref{sec:eff_one_body}.
Table~\ref{tab:mean_field} summarizes the mean-field decoupling in the TCKL.

\begin{table}[t]
\begin{tabular}{c|c}
\hline
\  Order Param. \  & \  Mean Field $(V_1, W_1; V_2, W_2)$ \ 
\\ 
\hline
$\la \Psi^z \ra$ & $(\sqrt 2 V, 0; 0, 0)$, $(0, \sqrt 2 V; 0, 0)$
\\
$\la \Psi^x \ra$ & $(V, 0; V, 0)$, $(0, V; 0, V)$
\\
$\la \Phi^x \ra$ & $(V, 0; 0, V)$, $(0, V; V, 0)$
\\
\hline
\end{tabular}
\caption{
Examples of the choice for mean-field decoupling in the original TCKL Hamiltonian.
}
\label{tab:mean_field}
\end{table}

We explain why the combination of the diagonal and offdiagonal mean-fields $V$ and $W$ is necessary for the present pairing state.
Equation \eqref{eq:mean_field2} shows that $W$ represents the pairing between conduction electrons and localized pseudofermions.
However, if we had only $W$, there would be {\it no pairing among conduction electrons} as illustrated in the upper panel of Fig.~\ref{fig:diagram}(b).
In this case the ordered state is identical to the composite channel order with $\bm \Psi$.
Only when both $V$ and $W$ are finite, the pairing among conduction electrons appears as shown in the lower part of Fig.~\ref{fig:diagram}(b).

The mean-field theory in this subsection can be technically applied to all the temperature range.
However, here again we emphasize that the present mean-field description is justified at sufficiently low temperature with $T\ll T_{\rm c}$, since the normal state of the TCKL is an incoherent metal \cite{jarrell96}, which cannot be described by a simple one-body Hamiltonian.
Note also that the Kondo energy scale is not properly incorporated in the mean-field theory, and the description is only qualitative.

\section{Electromagnetic Response}

\subsection{Linear response kernel}
For staggered ordered states on a tight-binding lattice, we cannot use the usual formulation for electromagnetic responses in continuum.
Therefore we include the electromagnetic fields as a Peierls phase following the literatures \cite{scalapino92, scalapino93, kostyrko94}.
The kinetic energy term in the presence of external fields is given by
\begin{align}
{\cal H}_{\rm kin} = -t \sum_{i\al\sg} \sum_{\nu=x,y,z} \epn^{\imu e A_\nu (\bm R_i)} c^\dg_{i+\delta_\nu, \al\sg}c_{i\al\sg} + {\rm h.c.}
,
\end{align}
where the vector potential $\bm A(\bm q)$ and electronic charge $e$ are introduced.
The vector $\delta_\nu$ points to the nearest neighbor site along $\nu=x,y,z$ direction.
The current operator is given by differentiating the Hamiltonian with respect to vector potential.
Neglecting contributions higher than second-order terms, we can write the current as $j^{\rm tot}_\nu = \sum_\nu K_{\nu\nu' } A_{\nu'}$.
The kernel is separated into paramagnetic and diamagnetic contributions as $K_{\nu\nu'} = K_{\nu\nu'}^{\rm para} + K_{\nu\nu'}^{\rm dia}$, each of which is given by
\begin{align}
K^{\rm para}_{\nu\nu'} (\bm q, \imu\nu_m) &= \frac{1}{N} \int_0^{1/T} \hspace{-2mm} \diff \tau \, \la j^{\rm para}_\nu (\bm q, \tau) j^{\rm para}_{\nu'} (-\bm q) \ra \, \epn^{\imu\nu_m\tau}
, \label{eq:para_kernel_def}\\
K^{\rm dia}_{\nu\nu'} (\bm q, \imu\nu_m) &= - \frac{e^2 \delta_{\nu\nu'}}{N} \sum_{\bm k\al\sg} \frac{\partial^2 \ep_{\bm k}}{\partial k_\nu^2}
\la c^\dg_{\bm k\al\sg} c_{\bm k\al\sg} \ra
, \label{eq:dia_kernel_def}
\end{align}
respectively.
Here the paramagnetic current operator is defined by
\begin{align}
j_\nu^{\rm para}(\bm q) &= -\imu e t \sum_{i\al\sg} 
\left(  c^\dg_{i \al\sg} c_{i+\delta_\nu, \al\sg}
-
 c^\dg_{i+\delta_\nu, \al\sg} c_{i \al\sg}
\right)
\epn^{-\imu \bm q\cdot \bm R_i}
\end{align}
with $\nu_m = 2\pi m T$ being the bosonic Matsubara frequency.
We have used the relation $\la c^\dg_{\bm k\al\sg} c_{\bm k'\al\sg} \ra = \delta_{\bm k\bm k'} \la c^\dg_{\bm k\al\sg} c_{\bm k\al\sg} \ra$ in Eq.~\eqref{eq:dia_kernel_def}.

Now we calculate the electromagnetic kernel in the effective model $\tilde {\cal H}_{\rm eff}$ defined in Eq.~\eqref{eq:one-body_hamilt_tilde}.
We neglect the vertex correction for the current-current correlation function.
A straightforward calculation gives
\begin{align}
&K_{\nu\nu'}^{\rm para} (\bm q, \imu\nu_m)
=  - \frac{4e^2}{N} \sum_{\bm kpp'} v_{\bm k\nu} v_{\bm k \nu'}
\nonumber \\
&\hspace{10mm} \times 
\left( -a_{\bm k+\frac{\bm q}{2}, p} a_{\bm k-\frac{\bm q}{2}, p'}
+ b_{\bm k+\frac{\bm q}{2}, p} b^*_{\bm k-\frac{\bm q}{2}, p'} \right)
\nonumber \\
&\hspace{10mm} \times 
 \frac{f(E_{\bm k+\frac{\bm q}{2}, p}) - f(E_{\bm k-\frac{\bm q}{2}, p'})}{\imu \nu_m - E_{\bm k+\frac{\bm q}{2}, p} + E_{\bm k-\frac{\bm q}{2}, p'}}
, \label{eq:paramag_kernel}
\\
&K_{\nu\nu'}^{\rm dia} (\bm q, \imu\nu_m) 
= 
-\frac{4e^2\delta_{\nu\nu'}}{N} \sum_{\bm kp}  \frac{\partial^2 \ep_{\bm k}}{\partial k_\nu^2}
a_{\bm kp} f(E_{\bm kp})
, \label{eq:diamag_kernel}
\end{align}
where the velocity is defined by $\bm v_{\bm k}={\partial \ep_{\bm k}}/{\partial \bm k}$.
The coefficients are given by Eqs.~\eqref{eq:coef_a} and \eqref{eq:coef_b}.
In the derivation of Eq.~\eqref{eq:paramag_kernel}, we have used the relation
\begin{align}
\bm v_{-\bm k-\bm Q} = \bm v_{\bm k}
,
\end{align}
which is characteristic for staggered pairing.
For uniform superconductors, in contrast, we use the relation $\bm v_{-\bm k} = - \bm v_{\bm k}$, which gives the additional minus sign.

\subsection{Meissner kernel}

The presence of the Meissner effect is determined by evaluating $K_{\nu\nu} (\bm q\rightarrow \bm 0, 0)$.
One can easily confirm $K_{\nu\nu} (\bm q\rightarrow \bm 0, 0) = 0$ for the zero mean-field limit $V/T\rightarrow 0$, meaning that there is no Meissner effect.
At sufficiently low temperatures with $T/V\rightarrow 0$, on the other hand, we can use the relations $f(\ep_{\bm k}) = \theta (-\ep_{\bm k})$, $f(E_{\bm k+})=0$ and $f(E_{\bm k-})=1$ for finite $V$.
We then derive the Meissner kernel as
\begin{align}
&
K_{\nu\nu} (\bm q\rightarrow \bm 0, 0) =
\frac{2e^2}{3} \int_{-\infty}^{\infty} \hspace{-2mm}
\diff \ep \tilde \rho_0 (\ep) \left[ -\delta (\ep) + L(2\sqrt 2 |V|, \ep) \right]
,
\end{align}
where we have defined the functions
\begin{align}
\tilde \rho_0 (\ep) &= \frac 1 N \sum_{\bm k} \bm v_{\bm k}^2 \delta (\ep - \ep_{\bm k})
,
\\
L(a, \ep) &= \frac{(\ep+\sqrt{\ep^2 + a^2})^3(-\ep + 3\sqrt{\ep^2 + a^2})}{a^2 (\ep^2 + a^2)^{3/2}} \, \theta (-\ep)
.
\end{align}
We can obtain the behaviors in two extreme cases using this expression.
For small $V$, where the relevant energy scale is much smaller than the band width, we replace the density of states $\tilde \rho_0 (\ep)$ by the value at the Fermi level.
The integration can then be performed, and we obtain
\begin{align}
K_{\nu\nu} (\bm q \rightarrow \bm 0, 0) = 0 
\ \ \  (V/t\rightarrow 0)
.
\end{align}
Namely, the Meissner kernel vanishes in the weak-coupling limit at zero temperature.
This property is in contrast to the ordinary BCS theory as discussed later.
On the other hand, the strong coupling limit, although it is not realistic, is instructive to check the sign of the Meissner kernel.
The explicit form is given by
\begin{align}
K_{\nu\nu} (\bm q \rightarrow \bm 0, 0) = - \frac{2e^2}{3} \tilde \rho_0 (0)
\ \ \  (V/t\rightarrow \infty)
.
\end{align}
This diamagnetic response means that the system shows the ordinary Meissner effect.
Here only the electronic band at the Fermi level
given by Eq.~\eqref{eq:energy1} contributes to the Meissner kernel.
This point clearly distinguishes the normal metallic band from the present itinerant band composed of the Bogoliubov particles.

\begin{figure}[t]
\begin{center}
\includegraphics[width=65mm]{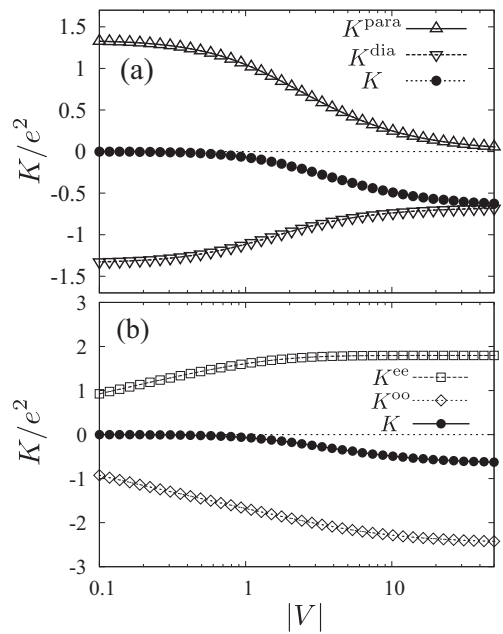}
\caption{
(a) Meissner kernel as a function of the mean-field $V$ $(\geq 0.1)$ at $T=0.005$ together with paramagnetic and diamagnetic kernels.
(b) Even- and Odd-frequency components of the Meissner kernel (see main text for definitions).
}
\label{fig:chi}
\end{center}
\end{figure}

The weak and strong coupling limits are smoothly connected as shown in Fig.~\ref{fig:chi},
where the contributions from paramagnetic and diamagnetic parts are also plotted separately.
Thus we have demonstrated that the positive (ordinary) Meissner effect is obtained in the OF superconductivity with staggered ordering vector.
The behavior is qualitatively same for the case away from half filling.

We comment on the characteristics of the electromagnetic response.
For small $V$, which is relevant to actual heavy-electron superconductors,
the Meissner kernel behaves as $K\propto - |V|^2$, and becomes zero in the zero mean-field limit.
This property is in contrast to the ordinary BCS-type pairing where the Meissner kernel is finite even in the weak-coupling limit (see Fig.~\ref{fig:other_chi}(b)).
Namely we have the ``weak'' Meissner effect in the present staggered OF superconductivity.
In analogy to the London equation in ordinary superconductors, the magnetic penetration depth is roughly given by $\lambda \sim 1/ \sqrt{-K}$.
The smaller Meissner kernel means the larger $\lambda$ compared to the BCS case.
The system then tends to be a type-II superconductor.

Now let us reconsider the Meissner kernel from a different point of view, focusing on frequency dependence of Green functions.
Since the kernel vanishes in the normal state, the existence of the anomalous Green function is essential for the Meissner effect.
Namely, the Meissner kernel can be rewritten as
\begin{align}
&K_{\nu\nu} (\bm q \rightarrow \bm 0, 0) 
= 
- \frac{8e^2T}{3} \sum_{\bm k n} \bm v_{\bm k}^2
{\cal F}_{\bm k}(\imu\ep_n) ^2
,
\end{align}
where we have chosen the phase as $\theta=0$ in the anomalous Green function ${\cal F}_{\bm k} (\imu\ep_n)$ defined in Eq.~\eqref{eq:def_f}.
For uniform OF superconductors, the anomalous Green function in the wave-vector space is purely odd with respect to frequency.
However, in the present pairing state with finite center-of-mass momentum $\bm Q$, the function ${\cal F}_{\bm k}(\imu\ep_n)$ is neither even nor odd function.
We then ask which part gives the dominant contribution to the Meissner kernel.

The anomalous Green function is decomposed into even and odd functions in frequency as 
${\cal F}_{\bm k}(\imu\ep_n) = {\cal F}^{\rm even}_{\bm k}(\imu\ep_n) + {\cal F}^{\rm odd}_{\bm k}(\imu \ep_n)$.
Correspondingly the kernel is written as $K=K^{\rm ee} + K^{\rm oo}$.
Note that cross terms such as $K^{\rm eo}$ become zero after taking the frequency summation.
As shown in Fig.~\ref{fig:chi}(b), $K^{\rm oo}$ gives the diamagnetic contribution, while $K^{\rm ee}$ gives paramagnetic contribution.
Hence the OF part of the anomalous Green function plays a dominant role for the positive Meissner effect in the present model.
In the ordinary BCS theory, on the other hand, the anomalous Green function in the wave-vector space has only the EF part, which gives the diamagnetic Meissner kernel.
Hence the result shown in Fig.~\ref{fig:chi}(b) is characteristic for the staggered pairing.
The difference lies in the relations $\bm v_{-\bm k}= - \bm v_{\bm k}$ for uniform pairing, and $\bm v_{-\bm k-\bm Q}= \bm v_{\bm k}$ for staggered pairing.

\subsection{Paramagnetic conductivity}
We consider the paramagnetic contribution for optical conductivity.
Putting $\bm q=\bm 0$ and $\imu\nu_m \rightarrow \omega+\imu \eta$ in Eq.~\eqref{eq:paramag_kernel}, we obtain
\begin{align}
\sigma^{\rm para}(\omega)&= 
{\imag K^{\rm para}_{\nu\nu} (\bm q=\bm 0, \omega+\imu\eta)}
/{\omega}
 \\
&= \frac{8\pi e^2 |V|^2}{3\omega^3} 
\tilde \rho_0 \hspace{-1mm} \left(
\frac{\omega^2 - 2 |V|^2}{\omega}
\right)
\theta \hspace{-1mm} \left( \omega - \sqrt 2 |V| \right)
, \label{eq:dynamical_kernel}
\end{align}
at $T=0$.
This expression shows that the optical conductivity has a gapped structure with the magnitude $\varDelta\omega = \sqrt 2 |V|$, even though we have a finite density of states at the Fermi level.
This is due to the fact that the single-particle gapless excitations are composed not of original electrons but of the Bogoliubov particles given by Eq.~\eqref{eq:bogolon_con}.
Hence there is no contribution from the intra-band transition which usually gives a Drude term.
Since the paramagnetic conductivity $\sigma^{\rm para}(0)$ is thus zero in the present pairing state,
the Drude weight in the total conductivity is the same as superfluid weight, which is identical to the Meissner kernel.



\subsection{Meissner kernel in related systems}

In this subsection, we explain the reason why we can have a positive Meissner effect in the staggered OF pairing state, by making a comparison with other types of pairings.
We examine the sign of the Meissner kernel both for EF and OF pairings taking the following simple models.

(a) {\it Uniform odd-frequency pairing.}---
It is possible to introduce a uniform OF pairing states, although it might not be stabilized thermodynamically.
Here we simply modify the staggered mean-field part in Eq.~\eqref{eq:one-body_hamilt_tilde} to the uniform one, and write down the Hamiltonian as
\begin{align}
{\cal H}^{\rm (a)} 
&= {\cal H}_{\rm kin}
+ |V| \sum_{\bm{k}\sg\sg'} 
\left[
 \epn^{\imu \theta/2} \delta_{\sg\sg'} f_{\bm k\sg}^\dg c_{\bm k 1 \sg}
 \right.
\nonumber \\
& \hspace{14mm} \left.
+ \,  \epn^{-\imu  \theta/2} \epsilon_{\sg\sg'} f_{\bm k\sg}^\dg c^\dg_{-\bm k, 2 \sg'}
+ {\rm h.c.}
\right] 
. \label{eq:hamilt_a}
\end{align}
The energy spectra are given by $E_{\bm k\pm}^{\rm (a)} = \pm \sqrt{\ep_{\bm k}^2 + 2|V|^2}$ and $E_{\bm k0}^{\rm (a)} = 0$.
The anomalous self energy of this model is given by
\begin{align}
\Delta_i^{(\rm a)} (\imu\ep_n) &= \frac{2 |V|^2 \epn^{\imu \theta}}{\imu\ep_n}
. \label{eq:self_ene_a}
\end{align}
The diagonal self energy $\Sigma_i^{(\rm a)} (\imu\ep_n)$ is the same as Eq.~\eqref{eq:self_ene1}.
Equation~\eqref{eq:self_ene_a} clearly shows the uniform OF superconductivity.

\begin{figure}[t]
\begin{center}
\includegraphics[width=85mm]{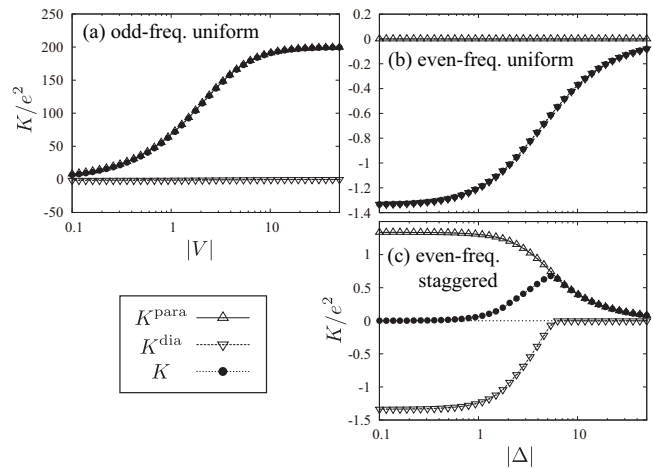}
\caption{
Meissner kernels for the Hamiltonians given by Eqs.~\eqref{eq:hamilt_a}, \eqref{eq:hamilt_b} and \eqref{eq:hamilt_c} as a function of mean-field at $T=0.005$.
}
\label{fig:other_chi}
\end{center}
\end{figure}

The Meissner kernel is calculated from Eqs.~\eqref{eq:para_kernel_def} and \eqref{eq:dia_kernel_def}.
Figure~\ref{fig:other_chi}(a) shows the numerical results for the mean-field dependence of the Meissner kernel together with paramagnetic and diamagnetic contributions.
The large value of the kernel is due to the presence of localized band at the Fermi level, which gives the contribution proportional to $1/T$ like a Curie law of localized spins in magnetism.
The sign of the Meissner kernel in Fig.~\ref{fig:other_chi}(a), which is our present concern, shows that the total response is paramagnetic.
Namely we have the negative Meissner effect in this model.
This result is what we have expected, as pointed out in the previous studies \cite{coleman94, heid95}.

(b) {\it Uniform even-frequency pairing.}---
The next example is the ordinary EF pairing, and the simplest is the $s$-wave spin-singlet pairing.
Here we just confirm the result in the BCS theory.
The model is given by the Hamiltonian
\begin{align}
{\cal H}^{(\rm b)} = {\cal H}_{\rm kin} + \sum_{\bm k \sg\sg'} \epsilon_{\sg\sg'}
\left( \Delta c^\dg_{\bm k\sg} c^\dg_{-\bm k\sg'} + {\rm h.c.} \right)
. \label{eq:hamilt_b}
\end{align}
Here we have considered only the spin degrees of freedom.
The anomalous self energy is given by
\begin{align}
\Delta_i^{(\rm b)} (\imu\ep_n) &= \Delta
,
\end{align}
and the energy spectra by $E_{\bm k\pm}^{\rm (b)} = \pm \sqrt{\ep_{\bm k}^2 + |\Delta|^2}$.
We show in Fig.~\ref{fig:other_chi}(b) the Meissner kernel.
The paramagnetic part becomes zero due to the gap structure in the single-particle spectrum, and only the diamagnetic part is finite.
Thus we have the positive Meissner effect in this case.

(c) {\it Staggered even-frequency pairing.}---
Finally we consider the staggered version of the EF pairing.
In a similar manner to Eq.~\eqref{eq:hamilt_b}, the corresponding Hamiltonian is simply written as
\begin{align}
{\cal H}^{(\rm c)} = {\cal H}_{\rm kin} + \sum_{\bm k \sg\sg'} \epsilon_{\sg\sg'}
\left( \Delta c^\dg_{\bm k\sg} c^\dg_{-\bm k-\bm Q, \sg'} + {\rm h.c.} \right)
. \label{eq:hamilt_c}
\end{align}
The single-particle energy spectrum and anomalous self energies are given by $E_{\bm k\pm}^{\rm (c)} = \ep_{\bm k} \pm |\Delta|$, and
\begin{align}
\Delta_i^{(\rm c)} (\imu\ep_n) &= \Delta \epn^{\imu \bm Q\cdot \bm R_i}
,
\end{align}
respectively.
We show in Fig.~\ref{fig:other_chi}(c) the mean-field dependence of the Meissner kernel.
The cusp structure at $|\Delta| = 6t$ corresponds to the band edge of the three-dimensional cubic lattice.
The total response is paramagnetic as shown in the figure, meaning that we have the negative Meissner effect.
Hence this staggered pairing state might not be thermodynamically stable.

\begin{table}[t]
\begin{tabular}{c|c|c}
\hline
 &\  uniform ($\bm q=\bm 0$)\  & \  staggered ($\bm q = \bm Q$) \  \\ 
\hline
\ EF pairing \ & \ \ positive\ \  &\ \  negative\ \  \\ 
\hline
\ OF pairing  \ & \ \ negative\ \  &\ \  positive\ \  \\ 
\hline
\end{tabular}
\caption{
Summary of properties of the Meissner kernel examined in this paper.
Here ``positive'' means the ordinary Meissner effect.
}
\label{tab:meissner}
\end{table}

We summarize in Tab.~\ref{tab:meissner} the results obtained in this subsection.
We conclude that both the OF property and staggered nature of the ordered state are important to make a positive Meissner effect in the present TCKL.

\section{Discussion}

Let us discuss characteristic properties of the staggered OF superconductivity in the TCKL.
One of them may be seen in the collective excitation mode.
As is well known, for ordinary uniform superconductors, the Nambu-Goldstone mode arising at $\bm q=\bm 0$ is absorbed into the gapped plasmon mode due to the long-ranged Coulomb interaction, and cannot be observed experimentally.
In contrast, the gapless mode appears at $\bm q=\bm Q$ for the staggered pairing state.
Hence the Nambu-Goldstone mode has a possibility to exist even in charged particle systems.
At present, this point remains to be clarified theoretically, which can be tested by incorporating the long-ranged Coulomb interaction explicitly in the model.
If the gapless excitation exists, the staggered superconductivity can be experimentally detected by observing the Nambu-Goldstone excitation mode in a charge dynamical structure factor at $\bm q = \bm Q$, since the phase fluctuation in pairing states is coupled to charge by uncertainty principle.
This low-energy mode will also affect thermodynamic quantities such as specific heat at low temperatures.

On the other hand, it is also characteristic that the staggered pairing state has a Fermi surface even in the superconducting state.
If this state persists to sufficiently low temperatures, the system shows the temperature-linear specific heat inside the pairing state.
However, this finite density of states at the Fermi level has a chance to cause fluctuations at low temperatures.
The TCKL in this case reaches the ground state in two steps; first the system becomes a Fermi liquid from a non-Fermi liquid by the gauge symmetry breaking, and then resolve the fluctuation originating from the Fermi surface by another phase transition.
Such possibility can be partly examined by using the mean-field theory formulated in this paper.

We finally comment on the staggered nature of the order parameter, which indicates that the ordering vector is dependent on the lattice geometry.
This motivates us to see what happens in the geometrically frustrated lattices.
If we consider the triangular lattice, for example, it is naively expected that a 120$^\circ$ N\'{e}el state for phase degrees of freedom is realized in analogy to the Heisenberg magnet.
In this case, the phase difference $\varDelta \theta = 2\pi/3$ between the nearest neighbor sites gives the finite current $I \propto \sin \varDelta \theta$, which results in the staggered loop current state \cite{tieleman13}.
Note that the staggered superconductivity in the present paper gives zero internal current because of the condition $\varDelta \theta =\pi$.
Another candidate state in triangular lattice is a stripe phase, where the phase difference is $\varDelta \theta = 0$ or $\pi$ which gives no internal current.
Thus it might be an interesting future problem to investigate the TCKL on geometrically frustrated lattices.

\section{Summary and Outlook} \label{sec_summary}

We have derived the low-energy fixed-point Hamiltonian for the staggered OF superconductivity in the TCKL.
Here the localized spins are effectively replaced by localized pseudofermions in the mean-field Hamiltonian.
The hybridization of conduction electrons with the pseudofermions plays an essential role for the formation of Cooper pairs among conduction electrons.
The constructed model reproduces the single-particle Green functions obtained by the DMFT in the lowest order in frequency.
The electromagnetic response function has also been microscopically calculated on the tight-binding lattice.
We have demonstrated that the ordinary Meissner effect is obtained in the present OF superconducting state.
Both the paramagnetic and diamagnetic parts give finite contributions to the Meissner kernel, and hence we have a small diamagnetic kernel in total.
By comparing the pairing state in the TCKL with the simple related models, we have clarified that the staggered nature of the order parameter plays an essential role for the positive Meissner effect.

The TCKL physically describes non-Kramers doublet systems in Pr or U-based compounds \cite{cox98}.
Hence the staggered OF superconductivity is possibly realized in actual TCKL systems, whose candidates include 
PrV$_2$Al$_{20}$ \cite{sakai12,matsubayashi12} and UBe$_{13}$ \cite{ott83,cox87} where strong $cf$ hybridization and unconventional superconductivity are found together with non-Fermi liquid behaviors.
The simple Hamiltonian \eqref{eq:hamilt}, however, is clearly insufficient to describe the details of real heavy-electron superconductors.
Complications beyond the simple model such as $f$-electron charge degrees of freedom, anisotropic exchange interaction and realistic conduction-electron band structures are necessary for this aspect, which will give us further insights on exotic superconductivities in actual compounds.


\section*{Acknowledgement}

The author is grateful to Y. Kato for valuable comments on the present paper, and to Y. Kuramoto and Y. Tanaka for fruitful discussions.
He also thanks M. Kunimi, N. Kurosawa and Y. Masaki for useful conversation on superconductivity.
This work is financially supported by the Japan Society for the Promotion of Science.

\end{document}